\documentclass[conference]{IEEEtran}
\IEEEoverridecommandlockouts
\usepackage{cite}
\usepackage{amsmath,amssymb,amsfonts}
\usepackage{algorithmic}
\usepackage{graphicx}
\usepackage{textcomp}
\usepackage{subcaption}
\usepackage[hyphens]{url} 
\usepackage[ruled,linesnumbered]{algorithm2e}
\usepackage{xcolor}
\def\BibTeX{{\rm B\kern-.05em{\sc i\kern-.025em b}\kern-.08em
    T\kern-.1667em\lower.7ex\hbox{E}\kern-.125emX}}
\begin{document}

\title{Knowledge-guided Deep Reinforcement Learning for Interactive Recommendation}
\author{Xiaocong Chen\textsuperscript{\rm 1}, Chaoran Huang\textsuperscript{\rm 1}, Lina Yao\textsuperscript{\rm 1}, Xianzhi Wang\textsuperscript{\rm 2}, Wei liu\textsuperscript{\rm 1},  Wenjie Zhang\textsuperscript{\rm 1}\\
\textsuperscript{\rm 1}School of Computer Science and Engineering, University of New South Wales, Australia\\ 
\textsuperscript{\rm 2}School of Computer Science, University of Technology Sydney, Australia \\
\textsuperscript{\rm 1}\{xiaocong.chen, chaoran.huang, lina.yao, wei.liu, wenjie.zhang\}@unsw.edu.au\\
\textsuperscript{\rm 2}xianzhi.wang@uts.edu.au\\
}
% \author{\IEEEauthorblockN{Xiaocong Chen}
% \IEEEauthorblockA{\textit{School of Computer Sci. \& Eng.} \\
% \textit{University of New South Wales}\\
% Sydney, Australia \\
% xiaocong.chen@unsw.edu.au}
% \and
% \IEEEauthorblockN{Chaoran Huang}
% \IEEEauthorblockA{\textit{School of Computer Sci. \& Eng.} \\
% \textit{University of New South Wales}\\
% Sydney, Australia  \\
% chaoran.huang@unsw.edu.au}
% \and
% \IEEEauthorblockN{Lina Yao}
% \IEEEauthorblockA{\textit{School of Computer Sci. \& Eng.} \\
% \textit{name of organization (of Aff.)}\\
% Sydney, Australia  \\
% lina.yao@unsw.edu.au}
% \and
% \IEEEauthorblockN{Xianzhi Wang}
% \IEEEauthorblockA{\textit{dept. name of organization (of Aff.)} \\
% \textit{name of organization (of Aff.)}\\
% Sydney, Australia  \\
% email address or ORCID}
% \and
% \IEEEauthorblockN{Wei Liu}
% \IEEEauthorblockA{\textit{School of Computer Sci. \& Eng.} \\
% \textit{name of organization (of Aff.)}\\
% Sydney, Australia  \\
% wei.liu@unsw.edu.au}
% \and
% \IEEEauthorblockN{Wenjie Zhang}
% \IEEEauthorblockA{\textit{School of Computer Sci. \& Eng.} \\
% \textit{name of organization (of Aff.)}\\
% Sydney, Australia  \\
% wenjie.zhang@unsw.edu.au}
% }

\maketitle

\begin{abstract}
Interactive recommendation aims to learn from dynamic interactions between items and users to achieve responsiveness and accuracy. 
Reinforcement learning is inherently advantageous for coping with dynamic environments and thus has attracted increasing attention in interactive recommendation research.
Inspired by knowledge-aware recommendation, we proposed Knowledge-Guided deep Reinforcement learning (KGRL) to harness the advantages of both reinforcement learning and knowledge graphs for interactive recommendation.
This model is implemented upon the actor-critic network framework.
It maintains a local knowledge network to guide decision-making and employs the attention mechanism to capture long-term semantics between items.
We have conducted comprehensive experiments in a simulated online environment with six public real-world datasets and demonstrated the superiority of our model over several state-of-the-art methods.
\end{abstract}

\begin{IEEEkeywords}
Recommender System, Reinforcement Learning, Deep Neural Network
\end{IEEEkeywords}

\section{Introduction}
\label{sec:intro}

Recommendation systems have been widely used by industry giants such as Amazon, YouTube, and Netflix to identify relevant, personalized content from large information spaces.
Modern recommendation systems are facing severe pressures for coping with emerging new users, ever-changing pools of recommendation candidates, and context-dependent interests~\cite{zhang2019deep}. In contrast, traditional recommendation methods focus on modeling user's consistent preferences and may not reflect high dynamics in user interest and environments.
In such situations, interactive recommendation rises as an effective solution that incorporates dynamic recommendation processes to improve the recommendation performance.
An interactive recommendation system would recommend items to an individual user and then receive the feedback to adjust its policies during the iterations~\cite{zhao2013interactive}.
Many studies model interaction recommendation as a Multi-Armed Bandit (MAB) problem~\cite{wang2016learning,wang2017community,wang2018online}.
Such methods generally assume a user's preference is consistent during the recommendation and focus on the trade-off between immediate and future rewards.
Therefore, they face challenges for handling environments with dynamically changing user preference or interest.
Reinforcement learning (RL) is a promising approach to interactive recommendation.
Considerable efforts have shown the outstanding performance of RL methods in recommendation systems~\cite{dulac2015deep,zhao2018deep,zheng2018drn}, thanks to its ability to learn from user's instant feedback.
Given its potential to handle dynamic interactions, RL has been widely regarded to be a possible better solution for interactive recommendation. 
However, most existing RL techniques in interactive recommendation focus on the usefulness instead of performance. For example, Liu et al.~\cite{liu2019diversity} employ the RL to increase the recommendation diversity, but not focus on the efficacy.
The primary reason is the agent only provides limited and partial information, making it difficult to control the decision-making process properly.
Besides, interactive recommendation systems usually contain a large number of discrete candidate actions, leading to high time complexity and low accuracy of RL-based techniques. Moreover, all the Deep Q-Networks (DQN)-based work~\cite{zhao2018recommendations,zhao2018deep,chen2018stabilizing,yao2018collaborative} gets struggled with a large number of discrete actions because DQN contains a maximise operation, which considers all actions. When the size of action increasing, the maximise operation will come to extremely slow, or even get stuck.
The policy gradient based methods will get stuck in this case as well because it may converge in the local maximum instead of the global maximum.

Recently, knowledge-aware recommendation systems have become popular as the knowledge graph can transfer the relation to contextual information 
and boost the recommendation performance~\cite{xian2019reinforcement,zhang2019quaternion}. Inspired by the above research, we propose a framework named knowledge-guided deep reinforcement learning (KGRL) for interactive recommendation.
We use the actor-critic framework to formulate the whole process.
Specially, we design a knowledge graph to represent relations between items so that the recommendation system can make recommendations based on the relations, and the critic network employs the knowledge graph as the guideline to improve the performance.

The critic network is used to evaluate the performance of the actor so as to let the actor optimize itself to the correct direction.
%which is similar with the goal of knowledge graph.
Besides, we apply graph convolutional network (GCN) inside the critic network capture the high-level structural information inside the knowledge graph and Deep Deterministic Policy Gradients (DDPG) to train our model.
In summary, we make the following contributions in this work:
\begin{itemize}
    \item We proposed a novel model KGRL where the knowledge graph is introduced into the reinforcement learning process to help the agent make decisions. 
    \item To improve the efficiency, we maintain a local knowledge network which is based on the knowledge graph, to fasten the process while keeping the performance;
    \item Comprehensive experiments in the simulated online environment with six real-world datasets prove the performance of our propose approach.
\end{itemize}

\section{Problem Definition}
\label{sec:problemDef}
An interactive recommendation system features incorporating user's feedback dynamically during the training process.
Given a set of users $\mathcal{U} = \{u, u_1, u_2, u_3, ...\}$ and a set of items $\mathcal{I} = \{i, i_1, i_2, i_3, ...\}$, the system first recommends item $i_1$ to user $u_1$ and then gets a feedback $x$.
The system aims to incorporate feedback to improve future recommendations.
To this end, it needs to figure out an optimal policy $\pi^*$ regarding which item to recommend to the user to achieve positive feedback.
We can formulate the problem as a Markov Decision Process (MDP) by treating the user as the environment and the system as the agent. We define the key components of the MDP as follows (Table~\ref{tab:symbols} summarizes the main notations used in this paper):
\begin{itemize}
    \item State: A state $S_t$ is determined by the recent $l$ items in which the user was interested before time $t$.
    \item Action: Action $a_t$ represents a user's dynamic preference at time $t$ as predicted by the agent.
    \item Reward: Once the agent chooses a suitable action $a_t$ based on the current state $S_t$ at time $t$, the user will receive the item recommended by the agent. The user's feedback on the recommended item (i.e., clicking the item, ignoring it) accounts for the reward $r(S_t,a_t)$, which will be considered to improve the recommendation policy $\pi$.
    \item Discount Factor $\gamma$: The discount factor $\gamma \in [0,1]$ is used to balance between the future and immediate rewards---the agent will fully focus on the immediate reward when $\gamma=0$ and take into account all the (immediate and future) rewards otherwise.
\end{itemize}

%list of symbols which be used in this study.
\begin{table}[t]
    \caption{Main notations}\smallskip
    \centering
    \begin{tabular}{c|l}
        \hline 
        Symbols & Meaning \\
        \hline 
         $\mathcal{U}$ & Set of users \\ 
         $\mathcal{I}$ & Set of items \\
         $\mathcal{R}$ & Set of relations \\
         $\mathcal{E}$ & Set of entities \\ 
         $|\cdot|$ & Number of unique elements in $\cdot$\\
         $\mathcal{S}_{u,t}$ & User $u$'s recent actions before timestamp $t$\\
         $\mathcal{G} = (\mathcal{E},\mathcal{R})$ & Constructed Knowledge Graph \\
         $\mathbb{E}$ & Item embedding \\
         $W$ & parameter matrices \\
         $S_t$ & state space at timestamp $t$\\
         $a_t$ & action space at timestamp $t$ \\
         $d$ & dimension of the latent space\\
         \hline
    \end{tabular}
    \label{tab:symbols}
\end{table}

\section{Methodology}
\label{sec:metho}
Our approach involves two steps: knowledge preparation and deep reinforcement recommendation
\subsection{Knowledge Preparation}
We construct the knowledge graph
%construction and transformation.  Commonly, the knowledge graph is constructed by 
based on entity-relation-entity tuples $\{(i,r,j) | i,j \in \mathcal{E}, r \in \mathcal{R}\}$. For example, the tuple \textit{(The Elements of Style, book.author, William Strunk Jr.)} means that \textit{William Strunk Jr} authored the book \textit{The Elements of Style}.
We consider every item (e.g., \textit{The Elements of Style}) as an entity in the knowledge graph $\mathcal{G}$ and transform the knowledge graph to represent user's preference more precisely~\cite{cao2019unifying}.
%To make the personalized recommendation, the transformation is necessary as the KG can not represent user's preference precisely~\cite{cao2019unifying}.
Given a user $u \in \mathcal{U}$ and an item $q \in \mathcal{I}$, suppose $\mathcal{D}(i)$ is the set of items that has direct relationship with item $i$ and $r_{ij}$ denotes the relation between items $i$ and $j$.
We calculate the user-specific relation scores as follows:
\begin{align*}
    f_u^{r_{ij}} = g(u,r_{ij}) \text{ where } g:\mathbb{R}^d \times \mathbb{R}^d \rightarrow \mathbb{R}
\end{align*}
where $g$ is a scoring function (e.g., inner product) to compute the score between user and relation; $d$ is the dimension of user representation and relation representation; $u \in \mathbb{R}^d, r_{ij} \in \mathbb{R}^d$;
$f_u^{r_{ij}}$ measures the importance of $r_{ij}$ to user $u$.

%For example, a user $u$ has already read the book ``The Elements of Style'', he/she may interested in another book which written by William Strunk Jr or a book with similar topic.
%All those potential candidates were involved in $\mathcal{D}(i)$ as those selection criteria are recognized as the direct relation.
Let $\mathcal{D}(i)$ be the set of candidates to recommend, we normalize the user-specific relation scores as follows:
%In order to make the scores (i.e., $f_u^{r_{ij}}$) in a same measure, the normalization technique was applied.
\begin{align*}
    \overline{f_u^{r_{ij}}} = \frac{f_u^{r_{id}}}{\sum_{d \in \mathcal{D}(i)} f_u^{r_{id}}} \in [0,1]
\end{align*}

\begin{figure*}[!ht]
    \centering
    \includegraphics[width=0.9\linewidth]{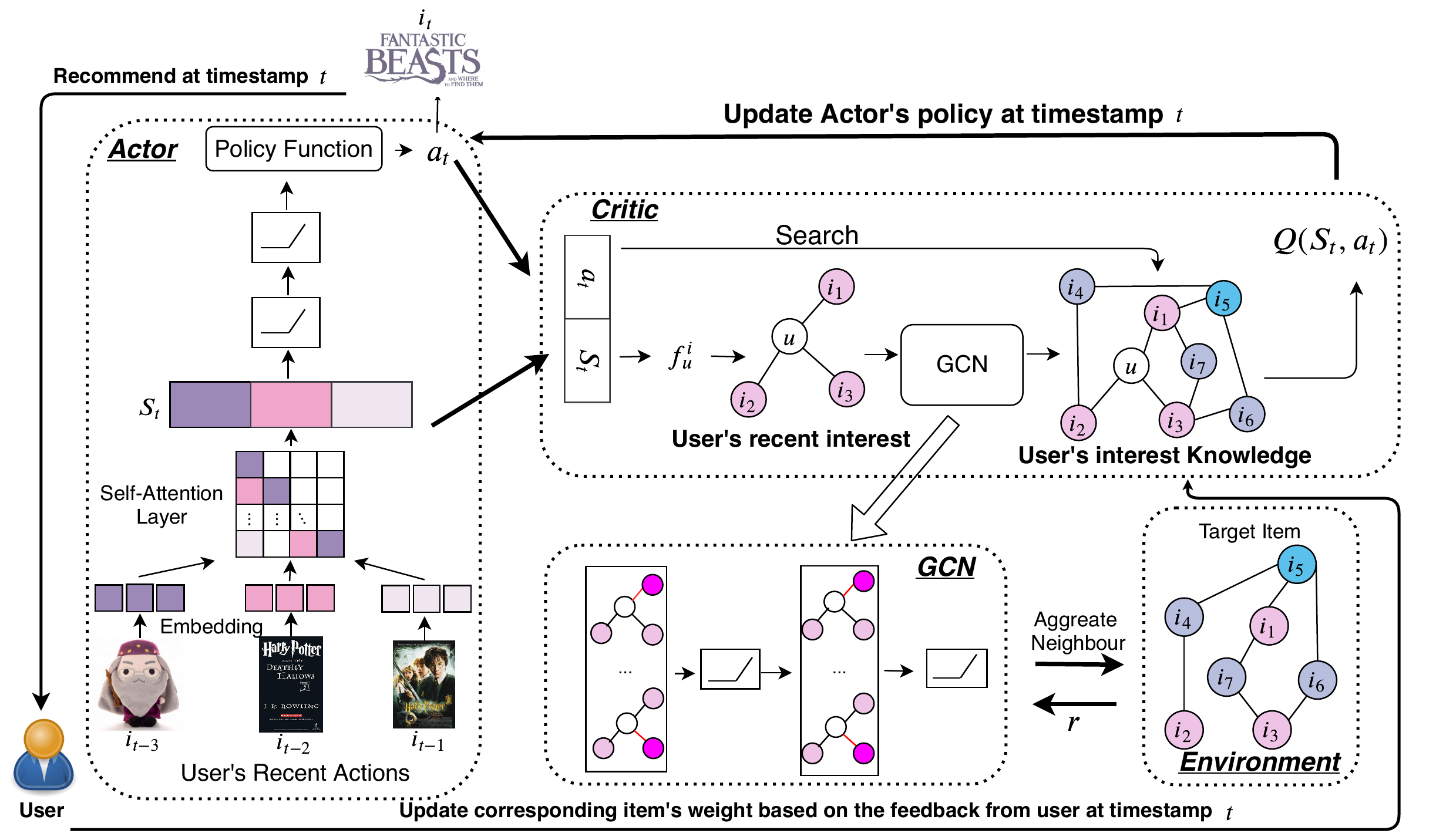}
    \caption{The KGRL structure.
    The left and right parts describe the actor network and the critic network, respectively, at time $t$.
    The model takes user's recent actions (regarding toys, books, and movies) as the input and recommends new items as the output. Those actions will be represented as the latent factor in this model.
    The user, in turn, provides feedback for the model to update user's interest knowledge's weights.}
    %The models takes user's recent actions (e.g., Dumbledore toy, Book of Harry Potter, and The movie of Harry Potter) as input . The model will recommend the movie ``Fantastic Beasts And Where To Find Them'' as they have same author. The user will need to provide the feedback which will be used to update the user's interest knowledge's weight (e.g., Yes or No for the recommended item).}
    \label{fig:structure}
\end{figure*}

Inspired by~\cite{wang2019knowledge}, we transform the knowledge graph into a user-specific graph $A_u$,
%using a user-specific relation scoring function $f_u^r$.
%The resulting graph is user-centered, with a user node in the center surrounded by nodes representing all relevant items.
%By given a normalized user score function $\overline{f}(\cdot)$, 
%Given normalized user scores, we can then convert the knowledge graph into a user-specific graph $A_u$.
which is an adjacency matrix of $\mathbb{R}^{|\mathcal{I}| \times |\mathcal{I}|}$.
In this matrix, each position $(i,j)$ corresponds to a score $\overline{f_u^{r_{ij}}}$,
%For instance, $A_u^{ij} = \overline{f_u^{r_{ij}}}$ which infers the value in matrix $A_u$ at position $(i,j)$ is $\overline{f_u^{r_{ij}}}$. 
and a higher score indicates a stronger relation between two items $i$ and $j$.
%Specially, a score of 0 means there no explicit relation between items $i$ and $j$.

\subsection{Deep Reinforced Recommendation}
We develop our recommendation model (Figure \ref{fig:structure}) based on the Actor-Critic reinforcement learning framework~\cite{grondman2012survey}, where the actor generates actions, %$a_t$
the critic evaluates actions, and the actor network updates the policy based on the suggestion made by the critic.

\subsubsection{Actor Network $\phi$} Given a current state $S_t$, the actor network employs a neural network to infer an optimal policy $\pi^*$ to work out an action $a_t$.
Given $S_t$, which consists of user's recent interests (shown in Figure \ref{fig:structure}, we first obtain vector representation of user's recent interest via embedding.
Suppose we have a set of user'srecently interested items before time $t$, $\mathcal{S}_{u,t} = \{\mathcal{S}_u^1,\mathcal{S}_u^2, ... ,\mathcal{S}_u^l\}$.
%which used to represent the recent $l$ items which user interested in . 
%We can describe our
The actor network
%as a network which 
takes an input sequence $\mathcal{S}_{u,t}$ and the corresponding feedback sequence
$\{\mathcal{F}_u^1,\mathcal{F}_u^2, ... ,\mathcal{F}_u^l\}$ to deliver
%the network should give 
an output sequence
$\{\mathcal{S}_u^2,\mathcal{S}_u^3, ... ,\mathcal{S}_u^{l+1}\}$.
%In the ideal situation, 
%Ideally, the user $u$ have a previous interested in item $i_1$, and provide the feedback $F_u^1$, the recommendation will recommend the item $i_2$.
%As time goes by, the number of previous items will increase, we need to insure the order of previous items is fixed.
Given an original item embedding matrix $\mathcal{M} \in
\mathbb{R}^{|\mathcal{I}| \times d}$ ($d$ is the dimension of the latent space), we apply positional embedding~\cite{kang2018self}, $P \in \mathbb{R}^{n \times l}$, to preserve the order of user's previously interested items, which updates the item embedding into the following:
%to achieve this goal.
%The positional embedding is learnable defined as $P \in \mathbb{R}^{n \times l}$, so the item embedding comes to:}
\begin{align*}
    \mathbb{E}=\begin{bmatrix}
     \mathcal{M}_1 + P_1   \\
     \mathcal{M}_2 + P_2\\
     \hdots \\
     \mathcal{M}_l + P_l\\
    \end{bmatrix}
\end{align*}

%After we get the item embedding,
We then fed this embedding into a self-attention layer to reduce impurity in the embedding~\cite{zhou2018atrank}.
The layer uses the scaled dot-product attention~\cite{vaswani2017attention}, which is originally defined as follows:
\begin{align*}
    \text{Attention}(Q,K,V) = \text{softmax}(\frac{QK^T}{\sqrt{d_k}})V 
\end{align*}
where $Q,K,V$ denotes queries, keys, and values, respectively; $\sqrt{d_k}$ is the scaling factor to regulate the value range of $QK^T$.
After applying the embedding $\mathbb{E}$ as the input, the attention turns into the following:
%of the attention layer is the embedding $\mathbb{E}$, the attention formula will comes to:
\begin{align*}
    \text{Attention}(\mathbb{E}W^Q, \mathbb{E}W^K, \mathbb{E}W^V)
\end{align*}
where $\mathbb{E}W^Q, \mathbb{E}W^K, \mathbb{E}W^V \in \mathbb{R}^{d \times d}$.
We fed this embedding into two fully connected layers, which use ReLU and tanh as the activation functions, respectively as described in~\cite{kang2018self}.
The output of the attention layer is the state $S_t$ at time $t$.

\subsubsection{Critic Network $\psi$} We design the critic network to estimate the Q-value function $Q(S_t,a_t)$ to evaluate actor's policy.
The critic network takes state representation $S_t$ and action representation $a_t$ as the input (shown in Figure \ref{fig:structure}).
We design a local knowledge network within the critic network to capture the high-order structural proximity among the items in the knowledge graph using graph convolutional network (GCN).
Specifically, given a user-specific graph $g_i^u$ generated from the current state $S_t$,
we feed it into a two-layer GCN that applies the following layer-wide propagation rule:
%To be specific, the layer-wide propagation rule can be written as follow~\cite{kipf2016semi}.
\begin{align}
    H^{l+1} = \sigma(D^{-\frac{1}{2}} \hat{A}_u D^{-\frac{1}{2}} H^l W^l)
\end{align}
where $H^{l+1}$ is the representation of entities at layer $l+1$; $A_u$ is the input matrix that aggregates the neighbour's entities; $\hat{A}_u$ is set to $A_u + I$, where the $I$ is an identity matrix used to avoid negligence of the old representation via self-connection; %can avoid negligence of the old representation;
$D_u$ is the diagonal degree matrix for 
$\hat{A}_u$ where $D_u^{ii} = \sum_j \hat{A}_u^{ij}$ (the symmetric normalization was applied to keep the representation $H^l$ stable, as denoted by $D^{-\frac{1}{2}} \hat{A}_u D^{-\frac{1}{2}}$); $W^l$ is the weight matrix for layer $l$; and $\sigma(\cdot)$ denotes the  non-linear activation function. 

Recent research shows the feasibility of searching in graphs processed by GCN~\cite{li2018combinatorial}.
Since GCN capture's all the structural information in the knowledge graph, it will not affect the search results.
In this study, we assume an unweighted graph where a user is equally interested in every item.
Then, we start searching with the actor predicted action $a_t$ (i.e., predicted item $i_p$) to the real target $i_t$, based on the user's personalized interest knowledge (i.e., trained graph with all parameters $\theta_{kg}$).
%we can start 
Finally, we calculate the Q value by estimating the reward $r$ based on the distance between the predicted item and the target item:
%and the Q value will be calculated to approximate this reward.
\begin{align*}
    r = \frac{100}{\sqrt{\text{Distance}(i_p,i_t) + \epsilon}} * W_{pt}
\end{align*}
where $W_{pt}$ is the sum of weight of the shortest path from $i_p$ to $i_t$; $\epsilon$ is the parameter to avoid the denominator becoming 0. We calculate the distance using the Dijkstra's algorithm with MinHeap.
%The .
%Furthermore, the user'sfeedback will be used to update the corresponding weight in local knowledge graph.

\subsection{Complexity Analysis}
We analyze the time and space complexity of the critic network, especially the search part, in this section.
%Empirically,
We consider a vector composition (i.e., the combination of the state vector and action vector) and assume the transmission time as a constant $c$.
%The majority part we want to discuss is the search.
%As mentioned, we have presented the graph as an adjacency matrix
%and use the Dijkstra's algorithm with MinHeap for the search.
Given a user interested in $I_u$ items, we consider the worst case---a complete graph and each item $i$ having $M$ nearest non-duplicate neighbours.
%(in the worst case, the neighbours don't have duplicate). Therefore,
Thus, we get a graph with $I_u + I_uM$ nodes (exclude the centralised user node) and $(I_u + I_uM)(I_u+I_uM -1)/2$ edges.
We then calculate the time and space complexity as
$\mathcal{O}((|I_u+I_uM)^2 + |I_u+I_uM|\log|I_u+I_uM|) \sim \mathcal{O}(|I_u+I_uM|^2)$ and $\mathcal{O}(2|I_u+I_uM|) \sim \mathcal{O}(|I_u+I_uM|)$.
In comparison, if we feed the environment knowledge graph to the critic network directly,
the time and space complexity would be $\mathcal{O}(|I + IM|^2)$
and $\mathcal{O}(|I+IM|)$.
Apparently, the local knowledge network significantly improves the performance and saves the memory space in our model ($I_u \ll I$). Moreover, the local knowledge network is easier to converge as it has fewer nodes than the environment knowledge graph.

\subsection{Training Strategy}
Training the actor-critic network requires train two parts of the neural network simultaneously.
We apply the Deep Deterministic Policy Gradient (DDPG) (Algorithm 1) to train our model~\cite{lillicrap2015continuous}, where we train the critic by minimising a loss function:
\begin{align*}
    & l(\theta_\psi) = \frac{1}{N} \sum_{j=1}^N ((r + \gamma \xi)-\psi_{\theta_\psi}(S_t,a_t))^2 \\
    & \text{where } \xi = \psi_{\theta_\psi'}(S_{t+1},\phi_{\theta_\phi'}(S_{t+1}))
\end{align*}
where $\theta_\psi$ is the parameter in critic; $\theta_\phi$ is the parameter in actor; 
$N$ is the size of mini-batch from the replay buffer; $\psi_{\theta_\psi'}$ and $\phi_{\theta_\phi'}$ are the target critic and target actor network, respectively.

Algorithm 2 describes the training of the local knowledge network,
%can be found in %\footnote{n-th order neighbours is the n-th nearest neighbours for item $i$ }.
where we define the same loss function for all users for the local knowledge network :
\begin{align*}
    l_k = \sum_{u \in \mathcal{U}} (\sum_{i:y_{ui}} J(y_{ui},\hat{y}_{ui}))
\end{align*}
where $J$ is the cross-entropy; $y_{ui}$ is a piece-wise function to reflect the interest/action (defined below):
\[ y_{ui}=\begin{cases} 
      1 & \text{if }$u$ \text{ interested in } $i$\\ 
      0 & \text{otherwise} 
   \end{cases}
\]
\begin{algorithm}[t]
\SetAlgoLined
 Initialize actor network $\phi$ with parameter $\theta_\phi$ and critic network $\psi$ with parameter $\theta_\psi$ randomly\; 
 Initialize target network $\phi'$ and $\psi'$\ with weight $\theta_\phi' \leftarrow \theta_\phi$, $\theta_\psi' \leftarrow \theta_\psi$ \;
 Initialize the local knowledge network \;
 Initialize Replay Buffer $\mathcal{B}$ \;
 \For{$i= 0$ to $n$}{
  Receive the initial state $S_i$ \;
  \For{$t=1$ to $T$}{
    Infer a action $a_t$ according to the $\phi(\cdot)$ \;
    Execute the action $a_t$ to receive a reward $r_t$ and observe a new state $S_{t+1}$\;
    $\mathcal{B}$.append($S_t, a_t, r_t, S_{t+1}$) \;
    Sample a random mini-batch of $\mathcal{N}$ transitions $(S_k, a_k, r_k, S_{k+1})$ from $\mathcal{B}$ \;
    Set $y_i = r_t + \gamma\xi$ \;
    Update Critic by minimise the loss $l(\theta_\psi)$ \;
    Update local knowledge net by Algorithm \ref{alg:kgcn} \;
    Update the Actor policy by using the sampled policy gradient:\\
     $\nabla_{\theta_{\phi}} \phi = \frac{1}{N} \sum_{j=1}^N \nabla_a \psi(S_k,a)|_{a=\phi(S_k)}\nabla_{\theta_\phi}\phi(S_k)$ \;
     Update target network:\\
      $\theta_\phi' \leftarrow \tau \theta_\phi + (1-\tau)\theta_\phi' $\;
      $\theta_\psi' \leftarrow \tau \theta_\psi + (1-\tau)\theta_\psi' $\;
  }
 }
 \caption{DDPG algorithm for our model}
 \label{alg:ddpg}
\end{algorithm}

\begin{algorithm}[t]
\SetAlgoLined
\SetKwInOut{Input}{input}
\Input{The user specific graph $g_i^u$, environment KG $\mathcal{G}_e$}
Initialize the parameters for GCN $\theta$ \;
Initialize the depth of graph $d_g$ \;
Initialize the reward storage $P$\;
\For{$i$ in $g_i^u$}{
    Receive the reward $r$ from $\mathcal{G}_e$ \;
    P.append(r)\;
}
$r$ = min(P)\;
\While{GCN is not converge}{
     \If{$d_g < r$}{
        aggregate next level's neighbours into $g_i^u$ 
        $d_g \leftarrow d_g + 1$\;
    }
    Update the GCN and its corresponding $\theta$\;
}
\caption{Training the local knowledge network}
\label{alg:kgcn}
\end{algorithm}

\section{Experiments}
\label{sec:expr}

In this section, we report our experimental evaluation of our model in comparison with several state-of-the-art models using real-world datasets. 

\subsection{Datasets} 
We conducted experiments on six public real-world datasets (Table~\ref{tab:stat} shows the statistics).
All these datasets provide the necessary information for building the respective knowledge graphs.

\begin{table}[t]
    \centering
    \caption{Statistics of our experimental datasets}\smallskip
    \resizebox{\columnwidth}{!}{%
    \begin{tabular}{r|r|r|r}
    %\small
        \hline 
        Dataset & \# of users & \# of items & \# of interactions \\
        \hline
         Amazon CD & 75,258 & 64,443 & 3,749,004\\
         Librarything & 73,882 & 337,561 & 979,053\\
         Book-Crossing & 278,858 & 271,379  & 1,149,780 \\
         GoodReads & 808,749 &  1,561,465 & 225,394,930\\
         MovieLens-20M &  138,493  & 27,278 & 20,000,263 \\ 
         Netflix & 480,189  & 17,770 & 100,498,277 \\
         \hline
    \end{tabular}%
    }
    \label{tab:stat}
\end{table}
\vspace{0.5mm}\noindent{\textbf{Book-Crossing\footnote{\url{http://www2.informatik.uni-freiburg.de/~cziegler/BX/}}}}: This dataset contains user's demographic information and book information from the Book-Crossing community. It is extremely sparse with a density of 0.0041\%.

\vspace{0.5mm}\noindent{\textbf{MovieLens-20M\footnote{\url{https://grouplens.org/datasets/movielens/}}}}: This is a well-known benchmark dataset that contains 20 million ratings from around 140 thousand users on the MovieLens website. It also provides movie tags, which can be used to build relations in the knowledge graph.

\vspace{0.5mm}\noindent{\textbf{Librarything\footnote{\url{http://cseweb.ucsd.edu/~jmcauley/datasets.html}\#social\_data}}}: This dataset contains book review information collected from the librarything website.

\vspace{0.5mm}\noindent{\textbf{Amazon CDs and Vinyl\footnote{\url{http://jmcauley.ucsd.edu/data/amazon/}}}}: This is a highly sparse dataset that contains the product metadata, user reviews, ratings, and item relations, as part of the Amazon e-commence dataset.

\vspace{0.5mm}\noindent{\textbf{Netflix Prize\footnote{\url{https://www.kaggle.com/netflix-inc/netflix-prize-data}}}}: This dataset contains 100 million ratings from 480 thousand users and item information for yearly open competition to improve Netflix's recommendation performance.

\vspace{0.5mm}\noindent{\textbf{Goodreads\footnote{\url{http://cseweb.ucsd.edu/~jmcauley/datasets.html}\#goodreads}}}: This dataset contains user's ratings and reviews to books on the Goodreads book review website.

%Furthermore, there some datasets which the rating range is up to 10 like Book-Crossing, we .

\begin{figure*}[ht]
    \begin{subfigure}{0.33\linewidth}
        \includegraphics[width=\linewidth]{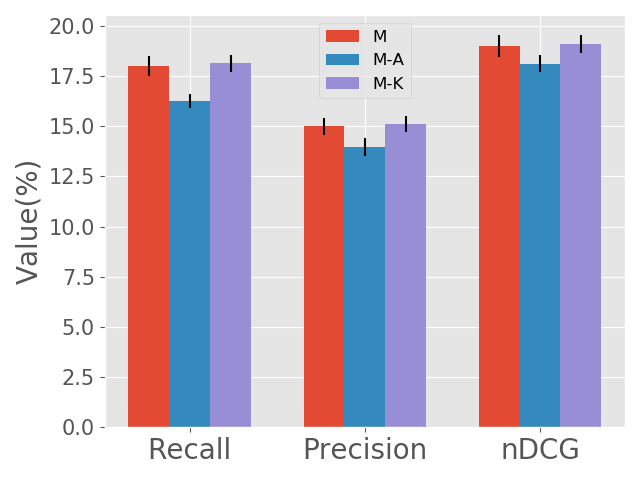}
        \caption{}
    \end{subfigure}
    \begin{subfigure}{0.33\linewidth}
        \includegraphics[width=\linewidth]{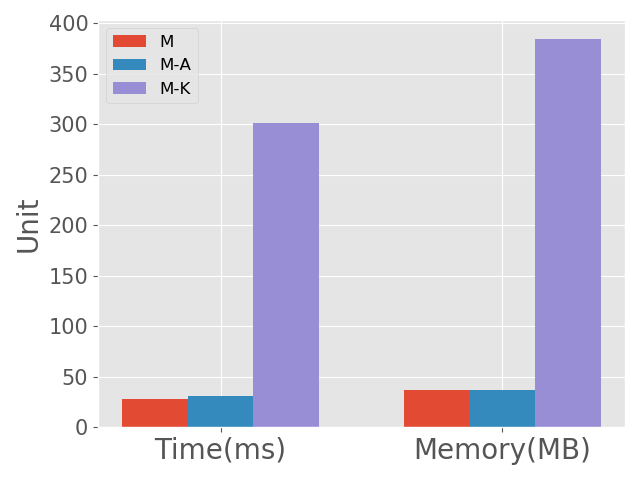}
        \caption{}
    \end{subfigure}
    \begin{subfigure}{0.33\linewidth}
        \includegraphics[width=\linewidth]{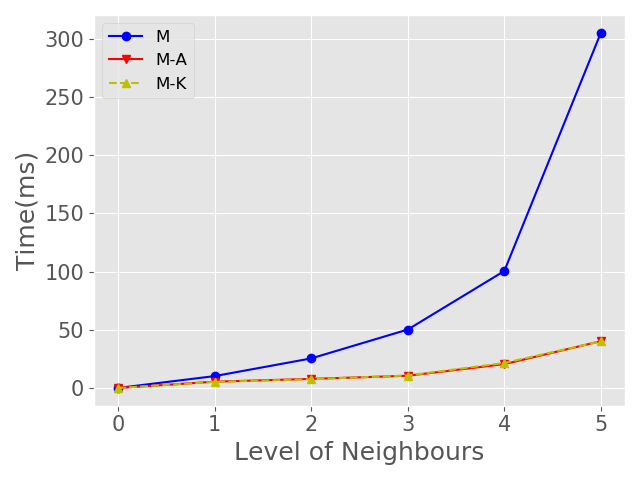}
        \caption{}
    \end{subfigure}
    \medskip
    \begin{subfigure}{0.33\linewidth}
        \includegraphics[width=\linewidth]{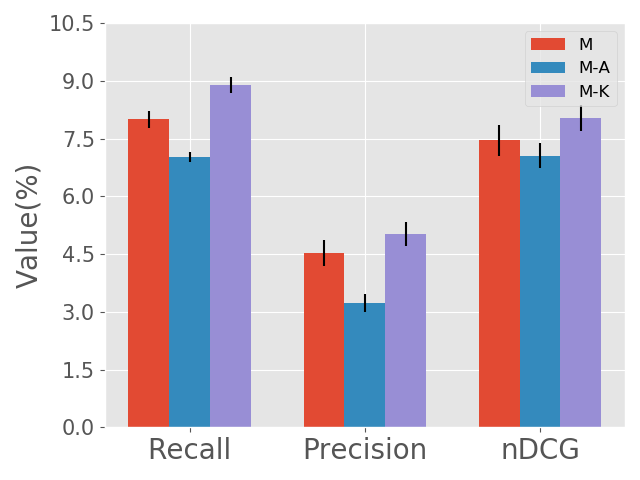}
        \caption{}
    \end{subfigure}
    \begin{subfigure}{0.33\linewidth}
        \includegraphics[width=\linewidth]{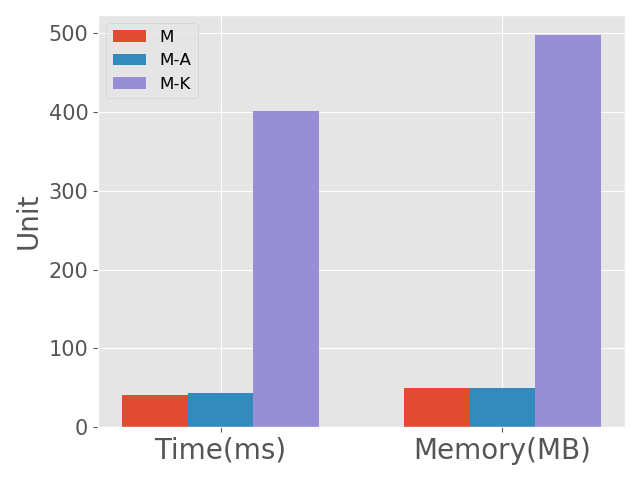}
        \caption{}
    \end{subfigure}
    \begin{subfigure}{0.33\linewidth}
        \includegraphics[width=\linewidth]{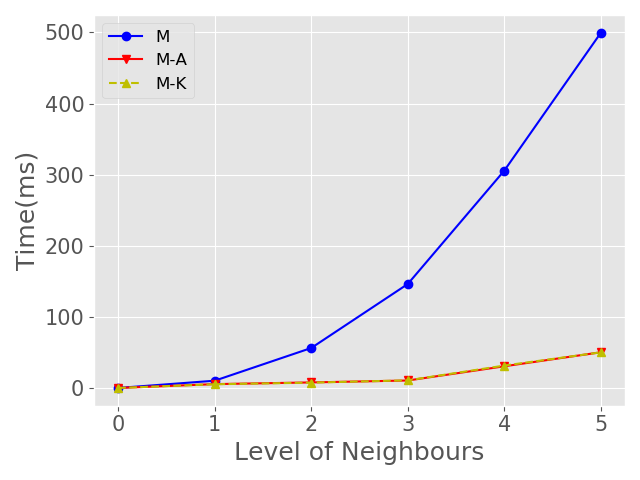}
        \caption{}
    \end{subfigure}
    \caption{Ablation and complexity studies on MovieLens-20M(a,b,c) and Book-Crossing(d,e,f): (a,d) Three models' performance in Recall, Precision, and nDCG; (b,e) Three models' time and memory consumption in conducting search for a target item located among fifth level neighbours; (c,f) Three models' time consumption along with an increasing level of the target item. \textit{M} denotes our original model, \textit{M-A} the model without the attention layer, meaning the item embedding will directly goes to state, and \textit{M-K} the model deprived of the local knowledge network---in this case, the model uses GCN to learn the whole environment inside itself.
    The level of neighbours represents the geographical location indicative of the shortest distance. For example, first-level neighbours represent the items which have a distance of 1 to the current item $i$.
    }
    \label{fig:aba}
\end{figure*}

\subsection{Evaluation Metrics}
We evaluate the performance of recommendation using three metrics: {precision}, {recall}, and {normalized Discounted
Cumulative Gain (nDCG)}.
All the metrics were calculated based on the top-10 recommendations to each user for each test case.
To ease processing, we removed users who have fewer than ten interactions and scaled the ratings from all datasets to the range of $[0,5]$.
Only the items with a rating score higher than three were considered a relevant item.
% For each user, the item with rating larger than 3 (the highest rating value is 5) are considered as the relevant item.

\subsection{Experimental Setup}
We evaluated our model in a simulated online environment built upon offline public datasets, using the algorithm proposed in~\cite{liu2019diversity} and the aforementioned reward function.
This way, we avoided collecting private user information and expensive online training~\cite{zhang2016collective}.
Specifically, the simulator generated feedback based on logistic matrix factorization (LMF)~\cite{johnson2014logistic}. We randomly split each dataset into a training set (70\%), a validation set (10\%), and a testing set (20\%) to conduct 10-fold cross-validation. The discount factor $\gamma$ was initialized to 0.99.

\subsection{Compared Methods}
We compared out model with several competitive baselines:

\vspace{0.5mm}\noindent{\textbf{Policy-Guided Path Reasoning (PGPR)~\cite{xian2019reinforcement}}}: A state-of-the-art knowledge-aware model that employs reinforcement learning for explainable recommendation.

\vspace{0.5mm}\noindent{\textbf{Tree-structured Policy Gradient Recommendation (TPGR)~\cite{chen2019large}}}: A state-of-the-art model that uses reinforcement learning and binary tree for large-scale interactive recommendation.
%in interactive recommendation. TPGR is designed to  problem by using the .
% MAB based follow

\vspace{0.5mm}\noindent{\textbf{HLinearUCB~\cite{wang2016learning}}}: A contextual-bandit approach that learns extra hidden features for each arm to model the reward for interactive recommendation.

\vspace{0.5mm}\noindent{\textbf{Wolpertinger~\cite{dulac2015deep}}}: A deep reinforcement learning framework that uses DDPG and KNN for recommendations in large discrete action spaces.

\vspace{0.5mm}\noindent{\textbf{DeepPage~\cite{zhao2018deep}}}: A DDPG-based reinforcement learning model that learns a ranking vector for page-wise recommendation.

\vspace{0.5mm}\noindent{\textbf{DRN~\cite{zheng2018drn}}}: A DQN-based recommendation method that employ deep Q learning to estimate Q-value for news recommendation.

\vspace{0.5mm}\noindent{\textbf{FactorUCB~\cite{wang2017factorization}}}: A matrix factorization-based bandit algorithm for interactive recommendation .

\vspace{0.5mm}\noindent{\textbf{ICTRUCB~\cite{wang2018online}}}: A MAB approach that uses a depend arm for online interactive collaborative filtering.

\subsection{Results}
Table \ref{tab:result} shows our evaluation results of recommendation models.
%by comparing with several state-of-art models, the fully result can be found in . From this table, we noticed that our model is
We observed that our model outperformed all the baselines in all metrics almost on all the datasets---it performed only slightly worse than TPGR on the Book-Crossing dataset.
This may be attributed to the specifical design of TPGR to deal with large-scale datasets.
%R is designed for large-scale dataset which may have a better ability to deal with very sparse dataset.
None of the MAB-based methods (HLinearUCB, FactorUCB and ICTRUCB) performed well on those datasets because they all assume static user interest and may not give up-to-date recommendations
%it may lead to the recommendation out-of-date.
We also observed that PGPR performed worse than DRN on the Amazon CD and Book-Crossing datasets---these sparse datasets might not provide sufficient relation for PGPR to infer the recommendation path.
%the PGPR is perform worse than the DRN. It may caused by the data sparsity because the 
Finally, all the models achieved their best results on the MovieLens-20M dataset, given the rich information and dense relation in the dataset.
%The possible explanation is that the distribution in MovieLens-20M is more dense than all the other dataset, after the pre-processing it may have more sufficient information for models to learn.
\begin{table*}[!ht]
\caption{The overall results of our model comparison with several state-of-arts models in different datasets. The result was reported by using the percentage and based on top-10 recommendation as mentioned before. The highlighted result in bold is the best result.}\smallskip
\begin{minipage}[ht]{1.0\linewidth}
\resizebox{\columnwidth}{!}{%
\begin{tabular}{ccccccc}
\hline
\multicolumn{1}{c|}{Dataset} & \multicolumn{3}{c|}{Amazon CD} & \multicolumn{3}{c}{Librarything} \\ \hline
\multicolumn{1}{c|}{Measure (\%)} & \multicolumn{1}{c|}{Recall} & \multicolumn{1}{c|}{Precision} & \multicolumn{1}{c|}{nDCG} & \multicolumn{1}{c|}{Recall} & \multicolumn{1}{c|}{Precision} & nDCG \\ \hline
\multicolumn{1}{c|}{Wolpertinger} & 1.542 $\pm$ 0.192 & 1.521 $\pm$ 0.145 & \multicolumn{1}{c|}{3.331 $\pm$ 0.201 } & 3.441 $\pm$ 0.313 & 3.673 $\pm$ 0.221 & 4.115$ \pm$ 0.251 \\ 
\multicolumn{1}{c|}{HLinearUCB} & 3.112 $\pm$ 0.331 & 2.647 $\pm$ 0.171 & \multicolumn{1}{c|}{4.005 $\pm$ 0.341} & 8.102 $\pm$ 0.396 & 7.431 $\pm$ 0.204 & 8.157 $\pm$ 0.241 \\ 
\multicolumn{1}{c|}{FactorUCB} & 3.531 $\pm$ 0.232 & 4.512 $\pm$ 0.242 & \multicolumn{1}{c|}{6.012 $\pm$ 0.251} & 8.541 $\pm$ 0.241 & 8.162 $\pm$ 0.355 & 8.653 $\pm$ 0.351   \\  
\multicolumn{1}{c|}{ICTRUCB} & 4.124 $\pm$ 0.293 & 3.110 $\pm$ 0.395 & \multicolumn{1}{c|}{5.982 $\pm$ 0.602} & 9.201 $\pm$ 0.241 & 7.980 $\pm$ 0.151 & 8.012 $\pm$ 0.466  \\  
\multicolumn{1}{c|}{DeepPage} & 7.124 $\pm$ 0.181 & 4.127 $\pm$ 0.134 & \multicolumn{1}{c|}{7.245 $\pm$ 0.154} & 10.342 $\pm$ 0.422 & 9.012 $\pm$ 0.241 & 9.124 $\pm$ 0.673 \\  
\multicolumn{1}{c|}{DRN} & 8.006 $\pm$ 0.232 & 4.234 $\pm$ 0.241 & \multicolumn{1}{c|}{6.112 $\pm$ 0.241} & 10.841 $\pm$ 0.112 & 9.412 $\pm$ 0.242 & 9.527 $\pm$ 0.455 \\  
\multicolumn{1}{c|}{TPGR} & 7.294 $\pm$ 0.312 & 2.872 $\pm$ 0.531 & \multicolumn{1}{c|}{6.128 $\pm$ 0.541} & 14.713 $\pm$ 0.644 & 12.410 $\pm$ 0.612 & 13.225 $\pm$ 0.722 \\  
\multicolumn{1}{c|}{PGPR} & 6.619 $\pm$ 0.123 & 1.892 $\pm$ 0.143 & \multicolumn{1}{c|}{5.970 $\pm$ 0.131 } & 11.531 $\pm$ 0.241 & 10.333 $\pm$ 0.341 & 12.641 $\pm$ 0.442  \\  
\hline
\multicolumn{1}{c|}{Ours} & \textbf{8.208 $\pm$ 0.241} & \textbf{4.782 $\pm$ 0.341} & \multicolumn{1}{c|}{\textbf{7.876 $\pm$ 0.511}} & \textbf{15.128 $\pm$ 0.241} & \textbf{12.451 $\pm$ 0.242} & \textbf{14.985$\pm$ 0.252} \\ 
\hline
\end{tabular}%
}
\end{minipage}

\begin{minipage}[ht]{1.0\linewidth}
\resizebox{\columnwidth}{!}{%
\begin{tabular}{ccccccc}
\hline
\multicolumn{1}{c|}{Dataset} &  \multicolumn{3}{c|}{Book-Crossing} & \multicolumn{3}{c}{GoodReads} \\ \hline
\multicolumn{1}{c|}{Measure (\%)} & \multicolumn{1}{c|}{Recall} & \multicolumn{1}{c|}{Precision} & \multicolumn{1}{c|}{nDCG} & \multicolumn{1}{c|}{Recall} & \multicolumn{1}{c|}{Precision} & nDCG \\ \hline
\multicolumn{1}{c|}{Wolpertinger}& 0.782 $\pm$ 0.121 & 1.235 $\pm$ 0.131 &\multicolumn{1}{c|}{0.976 $\pm$ 0.242} & 6.245 $\pm$ 0.122 & 3.415 $\pm$ 0.207 & 5.315 $\pm$ 0.321 \\ 
\multicolumn{1}{c|}{HLinearUCB}& 2.421 $\pm$ 0.131 & 1.724 $\pm$ 0.141 & \multicolumn{1}{c|}{2.865 $\pm$ 0.322} & 7.917 $\pm$ 0.303 & 5.151 $\pm$ 0.214 & 6.561 $\pm$ 0.351  \\ 
\multicolumn{1}{c|}{FactorUCB} & 3.123 $\pm$ 0.141 & 2.976 $\pm$ 0.223 & \multicolumn{1}{c|}{3.536 $\pm$ 0.241} & 5.643 $\pm$ 0.441 & 4.129 $\pm$ 0.221 & 6.122 $\pm$ 0.395 \\  
\multicolumn{1}{c|}{ICTRUCB} & 3.441 $\pm$ 0.121 & 3.421 $\pm$ 0.333 & \multicolumn{1}{c|}{4.001 $\pm$ 0.321} & 8.415 $\pm$ 0.132 & 6.432 $\pm$ 0.221 & 7.124 $\pm$ 0.241 \\  
\multicolumn{1}{c|}{DeepPage} & 5.124 $\pm$ 0.323 & 3.245 $\pm$ 0.142 & \multicolumn{1}{c|}{6.976 $\pm$ 0.142} & 10.071 $\pm$ 0.212 & 7.961 $\pm$ 0.232 & 8.329 $\pm$ 0.232 \\  
\multicolumn{1}{c|}{DRN} &  7.124 $\pm$ 0.122 & 4.123 $\pm$ 0.112 & \multicolumn{1}{c|}{7.433 $\pm$ 0.142 } & 10.620 $\pm$ 0.123 & 8.432 $\pm$ 0.241 & 9.461 $\pm$ 0.442 \\  
\multicolumn{1}{c|}{TPGR} & 7.246 $\pm$ 0.321 & \textbf{4.523 $\pm$ 0.442} & \multicolumn{1}{c|}{\textbf{7.870 $\pm$ 0.412}} & 13.219 $\pm$ 0.323 & 10.322 $\pm$ 0.442 & 9.825 $\pm$ 0.642 \\  
\multicolumn{1}{c|}{PGPR} & 6.998 $\pm$ 0.112 & 3.932 $\pm$ 0.121 & \multicolumn{1}{c|}{7.333 $\pm$ 0.133} &  11.421 $\pm$ 0.223 & 10.042 $\pm$ 0.212 & 9.234 $\pm$ 0.242 \\  
\hline
\multicolumn{1}{c|}{Ours} & \textbf{8.004 $\pm$ 0.223} & 4.521 $\pm$ 0.332& \multicolumn{1}{c|}{7.459 $\pm$ 0.401} & \textbf{13.444 $\pm$ 0.321} & \textbf{10.331 $\pm$ 0.331} & \textbf{11.641 $\pm$ 0.446}\\ 
\hline
\end{tabular}%
}
\end{minipage}

\begin{minipage}[ht]{1.0\linewidth}
\resizebox{\columnwidth}{!}{%
\begin{tabular}{ccccccc}
\hline
\multicolumn{1}{c|}{Dataset} &  \multicolumn{3}{c|}{MovieLens-20M} & \multicolumn{3}{c}{Netflix} \\ \hline
\multicolumn{1}{c|}{Measure (\%)} & \multicolumn{1}{c|}{Recall} & \multicolumn{1}{c|}{Precision} & \multicolumn{1}{c|}{nDCG} & \multicolumn{1}{c|}{Recall} & \multicolumn{1}{c|}{Precision} & nDCG \\ \hline
\multicolumn{1}{c|}{Wolpertinger} & 7.821 $\pm$ 0.171& 2.341 $\pm$ 0.142& \multicolumn{1}{c|}{4.002 $\pm$ 0.151} & 3.924 $\pm$ 0.222 & 2.911 $\pm$ 0.141  & 3.425 $\pm$ 0.261 \\ 
\multicolumn{1}{c|}{HLinearUCB}  & 13.591 $\pm$ 0.281 & 10.601 $\pm$ 0.132 & \multicolumn{1}{c|}{12.537 $\pm$ 0.285 } & 5.142 $\pm$ 0.314 & 5.052 $\pm$ 0.362 & 6.007 $\pm$ 0.425 \\ 
\multicolumn{1}{c|}{FactorUCB}  & 14.421 $\pm$ 0.412  & 11.229 $\pm$ 0.365 & \multicolumn{1}{c|}{11.422 $\pm$ 0.611 } & 5.643 $\pm$ 0.432  & 4.129 $\pm$ 0.233 & 6.122 $\pm$ 0.442 \\  
\multicolumn{1}{c|}{ICTRUCB}  & 14.345 $\pm$ 0.212 & 9.923 $\pm$ 0.222 & \multicolumn{1}{c|}{11.051 $\pm$ 0.423} & 7.00 1$\pm$ 0.312 & 6.212 $\pm$ 0.432 & 9.112 $\pm$ 0.523 \\  
\multicolumn{1}{c|}{DeepPage} & 12.472 $\pm$ 0.312 & 10.161 $\pm$ 0.332 & \multicolumn{1}{c|}{13.129 $\pm$ 0.322 } & 8.431 $\pm$ 0.212 & 7.324 $\pm$ 0.133 & 9.872 $\pm$ 0.223\\  
\multicolumn{1}{c|}{DRN}  & 14.742 $\pm$ 0.223 & 14.092 $\pm$ 0.342& \multicolumn{1}{c|}{16.245 $\pm$ 0.242} & 12.310 $\pm$ 0.144 & 10.213 $\pm$ 0.142 & 16.562 $\pm$ 0.153\\  
\multicolumn{1}{c|}{TPGR}  & 16.431 $\pm$ 0.369 & 13.421 $\pm$ 0.257 & \multicolumn{1}{c|}{18.512 $\pm$ 0.484} & 12.512 $\pm$ 0.556 & 11.512 $\pm$ 0.595 & 17.425 $\pm$ 0.602 \\  
\multicolumn{1}{c|}{PGPR} & 14.234 $\pm$ 0.207 & 9.531 $\pm$ 0.219 & \multicolumn{1}{c|}{11.561 $\pm$ 0.228} & 10.982 $\pm$ 0.181 & 10.123 $\pm$ 0.227 & 17.134 $\pm$ 0.243 \\  
\hline
\multicolumn{1}{c|}{Ours} & \textbf{18.021 $\pm$ 0.498} & \textbf{14.989 $\pm$ 0.432} & \multicolumn{1}{c|}{\textbf{19.007 $\pm$ 0.543}} & \textbf{13.009 $\pm$ 0.343} & \textbf{11.874 $\pm$ 0.232} & \textbf{19.082 $\pm$ 0.348} \\ 
\hline
\end{tabular}%
}
\end{minipage}
\label{tab:result}
\end{table*}

\subsection{Ablation and Complexity Studies}
We conducted ablation studies to explore the impact of the attention mechanism and local knowledge network on the performance of our model on the above six datasets. We selectively choose MovieLens-20M and the Book-Crossing as the example because the Book-Crossing dataset is the most sparse one and the MovieLens-20M is the most dense one; they can show the capability of our model in the normal case and extreme case. Due to the exponential increase in time usage, we only show the first five level of neighbours. 
%As presented in previous section, our model can spend much less time in the search process but keep the accuracy. We will examine this statement as well as the effect of the attention mechanism.
%model will have the whole environment KG inside and the GCN will be used to learn the whole environment)
%
The results (Figure \ref{fig:aba}(a,d)) show that our model's performance dropped slightly (by 1\% in precision, 2\% in recall, and 1\% in nDCG) without the attention mechanism while elevated slightly without the local knowledge network because the model already contains all the information, including abundant relation between items to support the decision making.
%without the local knowledge network have a slightly better result than our model is reasonable as it contain all the information which have abundant relation between items to support the decision making. However,

We also used valgrind\footnote{\url{http://www.valgrind.org/}} to monitor the memory usage, which, on the other hand, reveals the huge advantages of using a local knowledge network in reducing both the time and space complexity (also see Figure \ref{fig:aba}(b)). We mentioned that in figure \ref{fig:aba} (c,f), the model $M-K$ have an incredible increase in time consumption when the level goes over 2. One possible reason is that as the level goes higher, the graph comes more and more complex, which will affect the search critically.
\section{Related Work}
\label{sec:related_works}
%This work is closely related to and built upon three areas of research.

% \subsection{Interactive Recommendation}
Most existing work models interactive recommendation as a Multi-Armed Bandit (MAB) problem. And the primary solution lies in finding an Upper Confidence Bound (UCB).
Li et al.~\cite{li2010contextual} employ the first linear model to calculate the UCB for each arm.
Since then, many researchers combine other techniques such as matrix factorization, to find the UCB~\cite{wang2017factorization}.
For example, Wang et al.~\cite{wang2018online} proposed a new approach by choosing a dependent arm to calculate the UCB; Shen et al.~\cite{shen2018interactive}, instead, use deep learning-based methods to solve MAB. 

% \subsection{Reinforcement Learning based Recommendation}
Recent studies have shown the effectiveness of reinforcement learning in modeling interactions-related recommendation processes, where the recommendation problems are usually formulated as Markov Decision Processes.
One approach is based on Deep Q-learning (DQN)~\cite{mnih2015human}, which maximizes the Q-value from the predicted item and the target item.
Zheng et al.~\cite{zheng2018drn} combine the DQN with the Dueling Bandit Gradient Decent (DBGD)~\cite{grotov2016online} policy to recommend news.
Another thread of methods is DDPG-based~\cite{lillicrap2015continuous}.
Such methods aim to let the agent learn a proper policy instead of using the Q-value.
For example, Liu et al.~\cite{liu2019diversity} adopt DDPG to promote the diversity in interactive recommendation; Zhao et al.~\cite{zhao2018deep} use DDPG for page-wise recommendation.
% \subsection{Knowledge-aware Recommendation}
It is also worth mentioning that knowledge graphs can be useful for providing guidance in explainable recommendation~\cite{xian2019reinforcement}.
Knowledge-aware recommendation systems heavily rely on the use of relation inference to generate paths for recommendations~\cite{zhao2017meta}.
Wang et al.~\cite{wang2019kgn} show graph convolutional network can help learn neighbour representations and thus boost the recommendation performance. Another approach for knowledge aware recommendation is the embedding based~\cite{zhang2016collaborative,huang2018improving}.
\section{Conclusion and Future Work}
\label{sec:conc}
In this paper, we have proposed a knowledge-guided deep reinforcement learning framework (KGRL) for interactive recommendation.
KGRL uses the critic-actor learning framework to harness the interaction between users and the recommendation system and employs a local knowledge network to improve the stability and quality of the critic network for better decision-making.
%support the decision-making
%In order , we . 
Extensive experiments over an online simulator with six public real-world datasets demonstrate its superior performance over state-of-the-art models. To verify the effectiveness for each component, we conduct the ablation study for the local knowledge network and attention mechanism and selectively present the performance both in normal case and extreme case. 
% The ablation study manifest that the local knowledge network can significantly reduce the time and space complexity without dropping the accuracy.
%can acquire a better result.
We are planning to introduce various types of user information (e.g., user's thought when browsing items) to enrich the interaction and deploy our model in online business platforms to further test the performance in the future. In addition, the cold-start problem is another big challenge to be focused on. Besides, the algorithm \ref{alg:ddpg} used to train the model still lacks the knowledge about how to update step size will affect the training time and the convergence which can be solved in the future work.

\bibliographystyle{IEEEtran}
\bibliography{IEEEabrv,sample}

% Generated by IEEEtran.bst, version: 1.12 (2007/01/11)
\begin{thebibliography}{10}
\providecommand{\url}[1]{#1}
\csname url@samestyle\endcsname
\providecommand{\newblock}{\relax}
\providecommand{\bibinfo}[2]{#2}
\providecommand{\BIBentrySTDinterwordspacing}{\spaceskip=0pt\relax}
\providecommand{\BIBentryALTinterwordstretchfactor}{4}
\providecommand{\BIBentryALTinterwordspacing}{\spaceskip=\fontdimen2\font plus
\BIBentryALTinterwordstretchfactor\fontdimen3\font minus
  \fontdimen4\font\relax}
\providecommand{\BIBforeignlanguage}[2]{{%
\expandafter\ifx\csname l@#1\endcsname\relax
\typeout{** WARNING: IEEEtran.bst: No hyphenation pattern has been}%
\typeout{** loaded for the language `#1'. Using the pattern for}%
\typeout{** the default language instead.}%
\else
\language=\csname l@#1\endcsname
\fi
#2}}
\providecommand{\BIBdecl}{\relax}
\BIBdecl

\bibitem{zhang2019deep}
S.~Zhang, L.~Yao, A.~Sun, and Y.~Tay, ``Deep learning based recommender system:
  A survey and new perspectives,'' \emph{ACM Computing Surveys (CSUR)},
  vol.~52, no.~1, pp. 1--38, 2019.

\bibitem{zhao2013interactive}
X.~Zhao, W.~Zhang, and J.~Wang, ``Interactive collaborative filtering,'' in
  \emph{Proceedings of the 22nd ACM international conference on Information \&
  Knowledge Management}.\hskip 1em plus 0.5em minus 0.4em\relax ACM, 2013.

\bibitem{wang2016learning}
H.~Wang, Q.~Wu, and H.~Wang, ``Learning hidden features for contextual
  bandits,'' in \emph{Proceedings of the 25th ACM International on Conference
  on Information and Knowledge Management}.\hskip 1em plus 0.5em minus
  0.4em\relax ACM, 2016, pp. 1633--1642.

\bibitem{wang2017community}
X.~Wang, P.~Cui, J.~Wang, J.~Pei, W.~Zhu, and S.~Yang, ``Community preserving
  network embedding,'' in \emph{Thirty-First AAAI Conference on Artificial
  Intelligence}, 2017.

\bibitem{wang2018online}
Q.~Wang, C.~Zeng, W.~Zhou, T.~Li, S.~S. Iyengar, L.~Shwartz, and G.~Grabarnik,
  ``Online interactive collaborative filtering using multi-armed bandit with
  dependent arms,'' \emph{IEEE Transactions on Knowledge and Data Engineering},
  2018.

\bibitem{dulac2015deep}
G.~Dulac-Arnold, R.~Evans, H.~van Hasselt, P.~Sunehag, T.~Lillicrap, J.~Hunt,
  T.~Mann, T.~Weber, T.~Degris, and B.~Coppin, ``Deep reinforcement learning in
  large discrete action spaces,'' \emph{arXiv preprint arXiv:1512.07679}, 2015.

\bibitem{zhao2018deep}
X.~Zhao, L.~Xia, L.~Zhang, Z.~Ding, D.~Yin, and J.~Tang, ``a,'' in
  \emph{Proceedings of the 12th ACM Conference on Recommender Systems}.\hskip
  1em plus 0.5em minus 0.4em\relax ACM, 2018, pp. 95--103.

\bibitem{zheng2018drn}
G.~Zheng, F.~Zhang, Z.~Zheng, Y.~Xiang, N.~J. Yuan, X.~Xie, and Z.~Li, ``Drn: A
  deep reinforcement learning framework for news recommendation,'' in
  \emph{Proceedings of the 2018 World Wide Web Conference}.\hskip 1em plus
  0.5em minus 0.4em\relax IW3C2, 2018, pp. 167--176.

\bibitem{liu2019diversity}
Y.~Liu, Y.~Zhang, Q.~Wu, C.~Miao, L.~Cui, B.~Zhao, Y.~Zhao, and L.~Guan,
  ``Diversity-promoting deep reinforcement learning for interactive
  recommendation,'' \emph{arXiv preprint arXiv:1903.07826}, 2019.

\bibitem{zhao2018recommendations}
X.~Zhao, L.~Zhang, Z.~Ding, L.~Xia, J.~Tang, and D.~Yin, ``Recommendations with
  negative feedback via pairwise deep reinforcement learning,'' in
  \emph{Proceedings of the 24th ACM SIGKDD International Conference on
  Knowledge Discovery \& Data Mining}.\hskip 1em plus 0.5em minus 0.4em\relax
  ACM, 2018, pp. 1040--1048.

\bibitem{chen2018stabilizing}
S.-Y. Chen, Y.~Yu, Q.~Da, J.~Tan, H.-K. Huang, and H.-H. Tang, ``Stabilizing
  reinforcement learning in dynamic environment with application to online
  recommendation,'' in \emph{Proceedings of the 24th ACM SIGKDD International
  Conference on Knowledge Discovery \& Data Mining}.\hskip 1em plus 0.5em minus
  0.4em\relax ACM, 2018, pp. 1187--1196.

\bibitem{yao2018collaborative}
L.~Yao, Q.~Z. Sheng, X.~Wang, W.~E. Zhang, and Y.~Qin, ``Collaborative location
  recommendation by integrating multi-dimensional contextual information,''
  \emph{ACM Transactions on Internet Technology (TOIT)}, vol.~18, no.~3, pp.
  1--24, 2018.

\bibitem{xian2019reinforcement}
Y.~Xian, Z.~Fu, S.~Muthukrishnan, G.~de~Melo, and Y.~Zhang, ``Reinforcement
  knowledge graph reasoning for explainable recommendation,'' in
  \emph{Proceedings of the 42nd International ACM SIGIR Conference on Research
  and Development in Information Retrieval}.\hskip 1em plus 0.5em minus
  0.4em\relax ACM, 2019, pp. 285--294.

\bibitem{zhang2019quaternion}
S.~Zhang, Y.~Tay, L.~Yao, and Q.~Liu, ``Quaternion knowledge graph
  embeddings,'' in \emph{Advances in Neural Information Processing Systems},
  2019, pp. 2731--2741.

\bibitem{cao2019unifying}
Y.~Cao, X.~Wang, X.~He, Z.~Hu, and T.-S. Chua, ``Unifying knowledge graph
  learning and recommendation: Towards a better understanding of user
  preferences,'' in \emph{The World Wide Web Conference}.\hskip 1em plus 0.5em
  minus 0.4em\relax ACM, 2019, pp. 151--161.

\bibitem{wang2019knowledge}
H.~Wang, M.~Zhao, X.~Xie, W.~Li, and M.~Guo, ``Knowledge graph convolutional
  networks for recommender systems,'' in \emph{The World Wide Web
  Conference}.\hskip 1em plus 0.5em minus 0.4em\relax ACM, 2019, pp.
  3307--3313.

\bibitem{grondman2012survey}
I.~Grondman, L.~Busoniu, G.~A. Lopes, and R.~Babuska, ``A survey of
  actor-critic reinforcement learning: Standard and natural policy gradients,''
  \emph{IEEE Transactions on Systems, Man, and Cybernetics, Part C
  (Applications and Reviews)}, vol.~42, no.~6, pp. 1291--1307, 2012.

\bibitem{kang2018self}
W.-C. Kang and J.~McAuley, ``Self-attentive sequential recommendation,'' in
  \emph{2018 IEEE International Conference on Data Mining (ICDM)}.\hskip 1em
  plus 0.5em minus 0.4em\relax IEEE, 2018.

\bibitem{zhou2018atrank}
C.~Zhou, J.~Bai, J.~Song, X.~Liu, Z.~Zhao, X.~Chen, and J.~Gao, ``Atrank: An
  attention-based user behavior modeling framework for recommendation,'' in
  \emph{Thirty-Second AAAI Conference on Artificial Intelligence}, 2018.

\bibitem{vaswani2017attention}
A.~Vaswani, N.~Shazeer, N.~Parmar, J.~Uszkoreit, L.~Jones, A.~N. Gomez,
  {\L}.~Kaiser, and I.~Polosukhin, ``Attention is all you need,'' in
  \emph{Advances in neural information processing systems}, 2017, pp.
  5998--6008.

\bibitem{li2018combinatorial}
Z.~Li, Q.~Chen, and V.~Koltun, ``Combinatorial optimization with graph
  convolutional networks and guided tree search,'' in \emph{Advances in Neural
  Information Processing Systems}, 2018.

\bibitem{lillicrap2015continuous}
T.~P. Lillicrap, J.~J. Hunt, A.~Pritzel, N.~Heess, T.~Erez, Y.~Tassa,
  D.~Silver, and D.~Wierstra, ``Continuous control with deep reinforcement
  learning,'' \emph{arXiv preprint arXiv:1509.02971}, 2015.

\bibitem{zhang2016collective}
W.~Zhang, U.~Paquet, and K.~Hofmann, ``Collective noise contrastive estimation
  for policy transfer learning,'' in \emph{Thirtieth AAAI Conference on
  Artificial Intelligence}, 2016.

\bibitem{johnson2014logistic}
C.~C. Johnson, ``Logistic matrix factorization for implicit feedback data,''
  \emph{Advances in Neural Information Processing Systems}, vol.~27, 2014.

\bibitem{chen2019large}
H.~Chen, X.~Dai, H.~Cai, W.~Zhang, X.~Wang, R.~Tang, Y.~Zhang, and Y.~Yu,
  ``Large-scale interactive recommendation with tree-structured policy
  gradient,'' in \emph{Proceedings of the AAAI Conference on Artificial
  Intelligence}, vol.~33, 2019, pp. 3312--3320.

\bibitem{wang2017factorization}
H.~Wang, Q.~Wu, and H.~Wang, ``Factorization bandits for interactive
  recommendation,'' in \emph{Thirty-First AAAI Conference on Artificial
  Intelligence}, 2017.

\bibitem{li2010contextual}
L.~Li, W.~Chu, J.~Langford, and R.~E. Schapire, ``A contextual-bandit approach
  to personalized news article recommendation,'' in \emph{Proceedings of the
  19th international conference on World wide web}.\hskip 1em plus 0.5em minus
  0.4em\relax ACM, 2010, pp. 661--670.

\bibitem{shen2018interactive}
Y.~Shen, Y.~Deng, A.~Ray, and H.~Jin, ``Interactive recommendation via deep
  neural memory augmented contextual bandits,'' in \emph{Proceedings of the
  12th ACM Conference on Recommender Systems}.\hskip 1em plus 0.5em minus
  0.4em\relax ACM, 2018, pp. 122--130.

\bibitem{mnih2015human}
V.~Mnih, K.~Kavukcuoglu, D.~Silver, A.~A. Rusu, J.~Veness, M.~G. Bellemare,
  A.~Graves, M.~Riedmiller, A.~K. Fidjeland, G.~Ostrovski \emph{et~al.},
  ``Human-level control through deep reinforcement learning,'' \emph{Nature},
  vol. 518, no. 7540, p. 529, 2015.

\bibitem{grotov2016online}
A.~Grotov and M.~de~Rijke, ``Online learning to rank for information retrieval:
  Sigir 2016 tutorial,'' in \emph{Proceedings of the 39th International ACM
  SIGIR conference on Research and Development in Information Retrieval}.\hskip
  1em plus 0.5em minus 0.4em\relax ACM, 2016.

\bibitem{zhao2017meta}
H.~Zhao, Q.~Yao, J.~Li, Y.~Song, and D.~L. Lee, ``Meta-graph based
  recommendation fusion over heterogeneous information networks,'' in
  \emph{Proceedings of the 23rd ACM SIGKDD International Conference on
  Knowledge Discovery and Data Mining}.\hskip 1em plus 0.5em minus 0.4em\relax
  ACM, 2017, pp. 635--644.

\bibitem{wang2019kgn}
H.~Wang, F.~Zhang, M.~Zhang, J.~Leskovec, M.~Zhao, W.~Li, and Z.~Wang,
  ``Knowledge-aware graph neural networks with label smoothness regularization
  for recommender systems,'' in \emph{Proceedings of the 25th ACM SIGKDD
  International Conference on Knowledge Discovery \& Data Mining}.\hskip 1em
  plus 0.5em minus 0.4em\relax ACM, 2019.

\bibitem{zhang2016collaborative}
F.~Zhang, N.~J. Yuan, D.~Lian, X.~Xie, and W.-Y. Ma, ``Collaborative knowledge
  base embedding for recommender systems,'' in \emph{Proceedings of the 22nd
  ACM SIGKDD international conference on knowledge discovery and data
  mining}.\hskip 1em plus 0.5em minus 0.4em\relax ACM, 2016, pp. 353--362.

\bibitem{huang2018improving}
J.~Huang, W.~X. Zhao, H.~Dou, J.-R. Wen, and E.~Y. Chang, ``Improving
  sequential recommendation with knowledge-enhanced memory networks,'' in
  \emph{The 41st International ACM SIGIR Conference on Research \& Development
  in Information Retrieval}.\hskip 1em plus 0.5em minus 0.4em\relax ACM, 2018,
  pp. 505--514.

\end{thebibliography}

\end{document}